\documentclass[floatfix,nofootinbib,amsmath,amssymb,showpacs]{revtex4}

\usepackage[latin1]{inputenc}
\usepackage[T1]{fontenc}
\usepackage{graphicx}

\allowdisplaybreaks 

\begin{document}

\title{Infrared analysis of Dyson-Schwinger equations taking into account the Gribov horizon in Landau gauge}

\author{M. Q. Huber}
\email{markus.huber@uni-graz.at}
\affiliation{Institut f\"ur Physik, Karl-Franzens-Universit\"at Graz, Universit\"atsplatz 5, 8010 Graz, Austria}

\author{R. Alkofer}
\email{reinhard.alkofer@uni-graz.at}
\affiliation{Institut f\"ur Physik, Karl-Franzens-Universit\"at Graz, Universit\"atsplatz 5, 8010 Graz, Austria}

\author{S. P. Sorella}
\email{sorella@uerj.br}
\affiliation{
UERJ - Universidade do Estado do Rio de Janeiro, Instituto de F\'isica - Departamento de F\'isica Te\'orica, Rua S\~ao Francisco Xavier 524, 20550-013 Maracan\~a, 
Rio de Janeiro, Brasil
}

\date{\today}

\newcommand{\mhalfo}{\frac{1}{2}}	
\newcommand{\mhalf}[1]{\frac{#1}{2}}
\newcommand{\ka}{\kappa}
\newcommand{\equ}[1]{\begin{equation} #1 \end{equation}}
\newcommand{\ali}[1]{\begin{align} #1 \end{align}}
\newcommand{\eref}[1]{eq.~(\ref{#1})}
\newcommand{\fref}[1]{fig.~\ref{#1}}
\newcommand{\ddotp}[1]{\frac{d^d #1}{(2\pi)^d}}	
\newcommand{\nnnl}{\nonumber\\}	
\newcommand{\G}[1]{\Gamma(#1)}
\newcommand{\nq}{\nu_1}	
\newcommand{\nw}{\nu_2}	
\newcommand{\nd}{\nu_3}	
\newcommand{\de}{\delta} 
\newcommand{\dhalf}{\frac{d}{2}} 
\newcommand{\fig}[4]{\begin{figure}[#1]\centering\epsfig{file=#3}\caption{#4}\label{#2}\end{figure}}
\newcommand{\igfig}[4]{\begin{figure}[#1]\centering\includegraphics{#3}\caption{#4}\label{#2}\end{figure}}
\newcommand{\cb}{$\bigstar$} 
\newcommand{\ce}{$\blacksquare$} 

\begin{abstract}
The low momentum behavior of the Landau gauge Gribov-Zwanziger action is investigated using
the respective Dyson-Schwinger equations. Because of the mixing of the gluon and the auxiliary fields four scenarios can be distinguished for the infrared behavior. Two of them lead to inconsistencies and can be discarded. Another one corresponds to the case where the auxiliary fields behave exactly like the Faddeev-Popov ghosts and the same scaling relation as in standard Landau gauge, $\ka_A+2\ka_c=0$, is valid. Even the parameter $\ka$ is found to be the same, $0.595$. The mixed propagators, which appear, are suppressed in all loops, and their anomalous infrared exponent can also be determined. A fourth case provides an even stricter scaling relation that includes also the mixed propagators, but possesses the same qualitative feature, i.~e. the propagators of the Faddeev-Popov ghost and the auxiliary fields are infrared enhanced and the mixed and the gluon propagators are infrared suppressed. In this case the system of equations to obtain the parameter $\ka$ is non-linear in all variables.

\end{abstract}


\maketitle

\section{Introduction}

In recent years different methods have been used to determine Green functions, especially propagators, in Yang-Mills theory. A prominent choice of gauge is the Landau gauge due to its manifest Lorentz covariance and the minimum number of interaction terms, which renders it very accessible for functional methods. Whereas the ultraviolet parts can be obtained by perturbation theory, a more refined, i.~e. non-perturbative, treatment is necessary for the low and intermediate momentum regime. The asymptotic behavior at large distances is thereby of special interest, since the confinement scenarios of Kugo and Ojima \cite{Kugo:1979gm, Kugo:1995km} and Gribov and Zwanziger \cite{Gribov:1977wm,Zwanziger:1989mf,Zwanziger:1991gz,Zwanziger:1992qr} can be tested: The ghost propagator should be infrared (IR) enhanced, whereas the gluon propagator vanishes at zero momentum.

The original Gribov-Zwanziger scenario is based on an improved gauge fixing compared to the standard Faddeev-Popov theory \cite{Faddeev:1967fc}, which restricts the integration in field configuration space to the hyperplane $\partial_\mu A_\mu=0$. However, there are still configurations left that are related by gauge transformations, so that the Faddeev-Popov gauge fixing is insufficient. Gribov proposed by his no-pole condition a way to restrict integration to what is nowadays known as the Gribov region \cite{Gribov:1977wm}. Zwanziger generalized the no-pole condition to all orders leading to the horizon function \cite{Zwanziger:1989mf}, which is a non-local object. However, it can be localized by the introduction of more fields \cite{Zwanziger:1989mf}. An immediate consequence is that the perturbative gluon propagator changes its form due to a mixing at the level of two-point functions with the new fields: Instead of having a massless pole it vanishes at zero momentum. The quantization along Gribov-Zwanziger requires a new parameter, the Gribov parameter $\gamma$, whose value is fixed by a gap equation. Taking into account this equation, one can show that the Faddev-Popov ghost propagator gets IR enhanced like $1/k^4$ \cite{Gribov:1977wm,Zwanziger:1992qr,Gracey:2005cx}. Thus the improved restriction of integration in field configuration space alters the propagators significantly. This is a semi-perturbative result, since the perturbative propagators are considered but including the non-perturbative information the Gribov horizon provides. One should note that the UV part of the theory is not affected by the horizon function: In the asymptotic limit of very high momenta the propagators retain their usual form and also the renormalization of the theory requires no additional independent renormalization constants \cite{Zwanziger:1992qr,Maggiore:1993wq,Dudal:2005na}.

Of course it is interesting to employ a completely non-perturbative method to the investigation of propagators. Such a framework is provided by Dyson-Schwinger
(DSEs )\cite{Alkofer:2000wg} and renormalization group equations (RGEs) \cite{Berges:2000ew,Pawlowski:2005xe}. The former are the equations of motion of Green
functions and consist of an infinitely large tower of coupled equations. Therefore results in principle depend on the truncation. However, within the so-called
scaling solution it is possible to analyze consistently the qualitative behavior of \textit{all} Green functions in the IR. Again the 
gluon propagator in Landau gauge is found to be IR vanishing, going like $(p^2)^{2\ka-1}$, whereas the ghost gets IR enhanced as $(p^2)^{-\ka-1}$ \cite{vonSmekal:1997is,vonSmekal:1997vx}. The parameter $\ka$ is positive and can be evaluated as $0.595$ \cite{Lerche:2002ep,Zwanziger:2001kw}. The behavior found by Gribov and Zwanziger corresponds to $\ka=1$. One caveat of the use of Dyson-Schwinger equations up to now may be that only the standard Faddeev-Popov action has been used and Gribov copies were not taken into account explicitly. However, there exists an argument by Zwanziger, why the Dyson-Schwinger equations should not change, when derived with a proper restriction of the path integral \cite{Zwanziger:2001kw}. The key to this is that the determinant of the Faddeev-Popov operator, which appears in the path integral after the standard gauge fixing, vanishes at the first Gribov horizon. Thus, if one formally cuts the integration there, no additional surface terms are introduced and the DSEs are not altered.

The scaling type solution described above was the first solution found for both Landau gauge propagators using DSEs \cite{vonSmekal:1997is,vonSmekal:1997vx}. A few years ago another solution was obtained, called the decoupling solution, where the gluon propagator freezes to a constant value in the IR and the Faddeev-Popov ghost remains unenhanced \cite{Dudal:2007cw,Boucaud:2008ji,Aguilar:2008xm,Fischer:2008uz,Alkofer:2008jy}. This solution is favored by most recent lattice calculations \cite{Cucchieri:2007md,Cucchieri:2007rg,Cucchieri:2008fc,Bogolubsky:2009dc,Bornyakov:2008yx,Pawlowski:2009iv} and the refined Gribov-Zwanziger framework \cite{Dudal:2007cw,Dudal:2008sp} and can also be obtained by functional equations \cite{Boucaud:2008ji,Fischer:2008uz,Alkofer:2008jy}. The scaling solution on the other hand is found in lattice calculations at $\beta=0$ \cite{Sternbeck:2008na} and in two dimensions \cite{Maas:2007uv}, by DSEs and RGEs using appropriate boundary conditions \cite{vonSmekal:1997is,vonSmekal:1997vx,Fischer:2002hn,Pawlowski:2003hq,Fischer:2008uz,Alkofer:2008jy,Fischer:2009tn} and by stochastic quantization methods \cite{Zwanziger:2003cf,Zwanziger:2002ia,Zwanziger:2001kw}. So both solutions are found within functional methods, but it seems hard to get the scaling solution on the lattice. In this respect there has recently been a proposal \cite{Maas:2009se} on how to implement different boundary conditions as used in functional equations also on the lattice, so that both the family of decoupling solutions and the scaling solution can be obtained.

The use of a non-perturbative method might clarify the connection between the Gribov-Zwanziger and the standard Faddeev-Popov Lagrangians.
There have also been speculations that the negligence of Gribov copies in DSEs might be the reason for the observed discrepancy with lattice calculations. In this article we take a first step in this direction by analyzing the IR properties of the Gribov-Zwanziger action with Dyson-Schwinger equations. The basic techniques for such an endeavor are already known \cite{vonSmekal:1997vx,Alkofer:2004it,Fischer:2006vf,Huber:2007kc,Huber:2009wh}, but due to the mixing of the gluon field and the auxiliary fields some modifications are necessary. This mixing is indeed the main source of intricacies arising in this context: When the propagator of a field and its two-point function are no longer the inverse of each other, but there is a matrix relation involving several propagators/two-point functions, it is necessary to distinguish between their respective infrared exponents. We deal with the non-linear relation between the different dressing functions by splitting our analysis into four different cases, depending on which part is IR leading. We will show that two of these cases lead to inconsistencies so that only two possible solutions remain. In both of them the ghosts and the auxiliary fields have the same qualitative IR behavior. Their infrared exponent is related to that of the gluon propagator by the scaling relation $\ka_A+2\ka_{ghosts}=0$. The differences are in the details. The first solution has an independent mixed propagator. Since its contributions in DSEs are suppressed, the effective system of equations reduces to the same system as in standard Landau gauge. Consequently also the parameter $\ka:=\ka_{ghosts}\geq0$ is the same. The infrared exponent of the mixed two-point function is also calculated and shows that the mixed propagator is IR suppressed. The second solution provides also for the infrared exponent of the mixed propagator a scaling relation according to which it scales with the exponent $\ka/2-1$ and is thus IR suppressed.

Zwanziger found in a recent article an independent IR enhancement of the bosonic auxiliary field, where its propagator goes like $(p^2)^{-1-d/2}$ \cite{Zwanziger:2009je}. His ansatz differs from ours as he starts with the propagators, whereas we start with the two-point functions. Furthermore his solution is based on a cancelation in the determinant of the propagators that appears in the relation between the two-point functions and the propagators.

After setting the scene by presenting the usual Gribov-Zwanziger Lagrangian rewritten into a convenient form in sec. \ref{sec:DSEs}, we extend the known methods for power counting to the case of mixed propagators in sec. \ref{sec:PowerCounting}. The analysis is then carried out in sec. \ref{sec:Propagators} and we present our conclusions in Sec. \ref{sec:conclusions}. The appendix contains some details on the calculations of sec. \ref{sec:PowerCounting}.

\section{The Dyson-Schwinger equations for the Gribov-Zwanziger action}
\label{sec:DSEs}

The Lagrangian of Yang-Mills theory in Landau gauge is
\begin{align}
\mathcal{L}_{FP}&=\frac1{4}F^a_{\mu\nu}F^a_{\mu\nu}+\frac{1}{2\xi}(\partial_\mu A_\mu^a)^2-\bar{c}^a M^{ab}\, c^b,
\end{align}
with the field strength tensor
\begin{align}
F_{\mu \nu }^a&=\partial_\mu A_\nu^a-\partial_\nu A_\mu^a-g\,f^{abc}A_\mu^b A_\nu^c
\end{align}
and the Faddeev-Popov operator
\begin{align}
M^{ab}:=-\partial_\mu D^{ab}_\mu=-\delta^{ab}\Box-f^{abc}A^c_\mu \partial_\mu.
\end{align}
The fields $A$, $c$ and $\bar{c}$ denote the gluon, ghost and anti-ghost fields. $\xi$ is the gauge fixing parameter that is set to zero for Landau gauge, so that the integration in the path integral is restricted to the hyperplane $\partial_\mu A_\mu=0$ in field configuration space.
This restriction is sufficient, when only small fluctuations around $A=0$ occur (as in perturbation theory), since then no gauge equivalent configurations are taken into account. However, going to smaller momenta away from the ultraviolet, at some point so-called Gribov copies appear that falsify any calculations. These copies are gauge copies but within the hyperplane fixed by our choice of gauge. By restricting the field configuration space to the Gribov region, defined by those configurations that have only positive eigenvalues for the Faddeev-Popov operator \cite{Gribov:1977wm}, the problem gets alleviated, but still copies remain \cite{vanBaal:1991zw}. To single out a unique representative of a gauge orbit one can for example take those configurations along the gauge orbit that are in the global minimum with respect to their norm. On a lattice this is, however, a polynomially hard computational problem and therefore not feasible on large lattices. How a restriction of the integration region to unique field configurations could be implemented at the level of the Lagrangian is currently unknown. So we ignore these copies for now and will only talk about the Gribov region in the following, for which such a restriction can be done.

To accomplish this restriction one can add a non-local term to the Lagrangian called horizon function \cite{Zwanziger:1989mf}. This introduces a new variable
$\gamma$, the Gribov parameter. It is not free, but fixed by a gap equation. However, the non-local term can be localized using two pairs of auxiliary fields \cite{Zwanziger:1989mf}: The bosonic fields $\varphi$ and $\bar{\varphi}$ and the fermionic ones $\omega$ and $\bar{\omega}$. Here we give the localized expression, since we will only work with this form of the Gribov-Zwanziger Lagrangian:
\begin{align}\label{eq:GZ-localized}
\mathcal{L}_{GZ}=\bar{\varphi}_\mu^{ac}M^{ab}\varphi_\mu^{bc}-\bar{\omega}\mu^{ac}M^{ab}\omega_\mu^{bc}+\gamma^2 g\,f^{abc}A_\mu^a(\varphi_\mu^{bc}-\bar{\varphi}_\mu^{bc})-\gamma^4 d(N^2-1),
\end{align}
with the Gribov parameter $\gamma$ determined by
\begin{align}
\langle g\,f^{abc}A_\mu^a(\varphi_\mu^{bc}-\bar{\varphi}_\mu^{bc}) \rangle=2\gamma^2 \,d (N^2-1),
\end{align}
where $N$ is the number of colors and $d$ the dimension of space-time.

For our calculations we will split the bosonic ghost fields into real and imaginary parts $U$ and $V$, respectively, as done in ref. \cite{Zwanziger:2009je}:
\begin{align}
\mathcal{L}&=\mathcal{L}_{FP}+\mathcal{L}'_{GZ},\\
\mathcal{L}'_{GZ}&=\mathcal{L}_U+\mathcal{L}_V+\mathcal{L}_{UV}-\bar{\omega}M\omega,\\
\mathcal{L}_U&=\frac1{2} U_\mu^{ac}\,M^{ab}\,U_\mu^{bc},\\
\mathcal{L}_V&=\frac1{2} V_\mu^{ac}\,M^{ab}\,V_\mu^{bc} +i\, g\,\gamma^2 \sqrt{2} f^{abc} A_\mu^a V_\mu^{bc},\\
\mathcal{L}_{UV}&=\frac1{2}i\,g f^{abc} U_\mu^{ad} V_\mu^{bd} \partial_\nu A_\nu^c,
\end{align}
where $U$ and $V$ are defined by
\begin{align}
\varphi=\frac1{\sqrt{2}}\left( U+i\,V\right), \quad \bar{\varphi}=\frac1{\sqrt{2}}\left( U-i\,V\right).
\end{align}
When the Landau gauge condition $\partial_\mu A_\mu=0$ is enforced, $\mathcal{L}_{UV}$ vanishes and the only mixing on the level of two-point functions is between the gluon field $A$ and the imaginary part of the bosonic ghost field $V$, whereas the $U$-ghost does not mix. This splitting simplifies calculations, because we only have to deal with a two-by-two matrix instead of a three-by-three matrix for the mixing. Perturbative expressions for the propagators derived from this Lagrangian can be found in ref. \cite{Gracey:2009mj}.

Now we have three different types of fields (Faddeev-Popov ghosts $c$ and $\bar{c}$, fermionic auxiliary fields $\omega$ and $\bar{\omega}$, real part of the bosonic auxiliary field $U$) that only interact with the gluon field via the Faddeev-Popov operator. To alleviate calculations for our purposes it is useful that we can combine all these fields into one type only.
Since they appear quadratic in the action, they can be integrated out in the path integral:
\begin{align}
\int D[\bar{c}c] e^{\bar{c}\,M\,c}&=det\,M,\\
\int D[\bar{\omega}\omega] e^{\bar{\omega}\,M\,\omega}&=(det\,M)^{d(N^2-1)},\\
\int D[U] e^{-\frac1{2}U\,M \,U}&=(det\,M)^{-\frac{d}{2}(N^2-1)}.
\end{align}
The different exponents of the determinant of the Faddeev-Popov operator are due to the different numbers of degrees of freedom. For $\gamma=0$ also the $V$-field can be integrated out and all determinants from auxiliary fields cancel, so that the original Faddeev-Popov Lagrangian is recovered. For the purpose of this article we can treat all these non-mixing fields as new fermionic fields $\eta$ and $\bar{\eta}$ with the appropriate number of degrees of freedom, namely $\frac{d}{2}(N^2-1)+1$. The field $V$ cannot be included due to its mixing with the gluon field and therefore the two fields $V$ and $\eta$ can have a different infrared behavior. An overview of the different fields is given in tab. \ref{tab:ghosts}. The final Lagrangian reads
\begin{align}\label{eq:Lagrangian}
\mathcal{L}&=\frac1{4}F^a_{\mu\nu}F^a_{\mu\nu}+\frac{1}{2\xi}(\partial_\mu A_\mu)^2-\bar{\eta}^{a}_c \,M^{ab}\, \eta^{b}_c+\frac1{2} V_\mu^{ac}\,M^{ab}\,V_\mu^{bc} +i\, g\,\gamma^2 \sqrt{2} f^{abc} A_\mu^a V_\mu^{bc}+\frac1{2}i\,g f^{abc} U_\mu^{ad} V_\mu^{bd} \partial_\nu A_\nu^c,
\end{align}
where the subscript index of the new ghost fields $\eta$ and $\bar{\eta}$ runs from $1$ to $\frac{d}{2}(N^2-1)+1$. Note that for odd dimensions and even $N$ this number is half-integer and therefore this transformation is not directly possible. However, we can consider only integer values and perform an analytic continuation to half-integer values, if necessary. Alternatively one keeps the three ghosts separated and will get the appropriate numerical factors in front of the diagrams. A further diagonalization of the Lagrangian would require a further splitting of the fields in color space due to the different number of color indices of the $A$- and $V$-fields. The new fields would mean a significant complication of the Lagrangian at the level of vertices.
Therefore we continue with the expression given above.

\begin{table}[t]
\begin{tabular}{l|c|c}
field & \# degrees of freedom & statistics \\
\hline
\hline
$c$, $\bar{c}$ & $1$ & fermionic \\
$\omega$, $\bar{\omega}$ & $d(N^2-1)$ & fermionic \\
$\phi$, $\bar{\phi}$ & $d(N^2-1)$ & bosonic \\
\hline
$U$ & $d/2(N^2-1)$ & bosonic \\
$V$ & $d/2(N^2-1)$ & bosonic \\
$\eta$, $\bar{\eta}$ & $d/2(N^2-1)+1$ & fermionic
\end{tabular}
\caption{\label{tab:ghosts} The numbers of degrees of freedom and the statistics of the Faddeev-Popov ghosts $c$ and $\bar{c}$, the original auxiliary fields $\omega$, $\bar{\omega}$, $\phi$ and $\bar{\phi}$, the bosonic auxiliary fields $U$ and $V$ and the fermionic fields $\eta$ and $\bar{\eta}$.}
\end{table}

\begin{figure}
\begin{center}
\includegraphics[width=\textwidth]{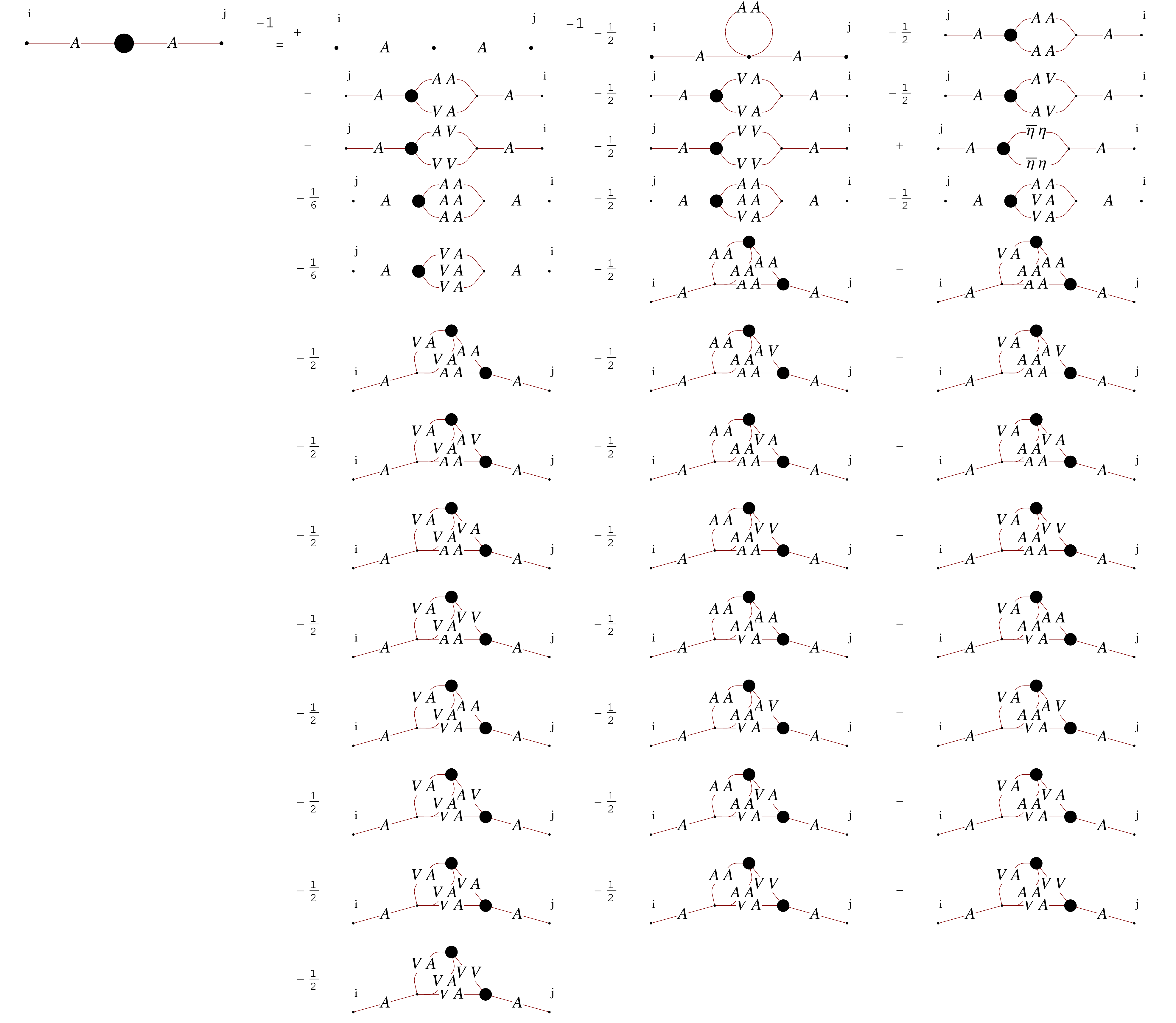}
\end{center}
\caption{\label{fig:AA-prop-DSEs-1L}The DSE of the gluon two-point function. A propagator with the exponent $-1$ denotes the two-point function.
This convention, strictly speaking being mathematically incorrect, is chosen for the purpose of diagrammatic representation only.\\
The propagators are labeled by the respective fields. The indices $i$ and $j$ denote the first and second fields of the depicted two-point function DSE.
}
\end{figure}

\begin{figure}
\begin{center}
\includegraphics[width=\textwidth]{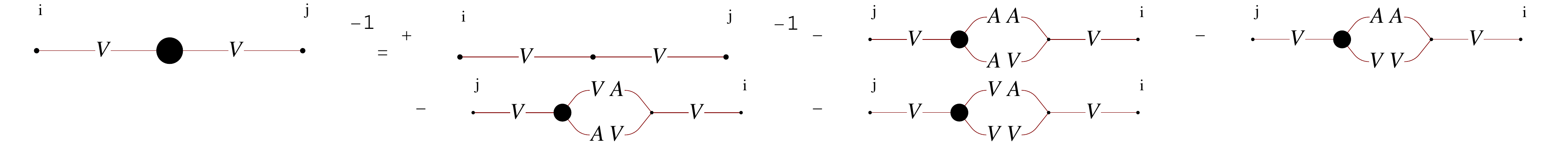}
\end{center}
\caption{\label{fig:VV-prop-DSEs-1L}The DSE of the $V$-field two-point function.}
\end{figure}

\begin{figure}
\begin{center}
\includegraphics[width=\textwidth]{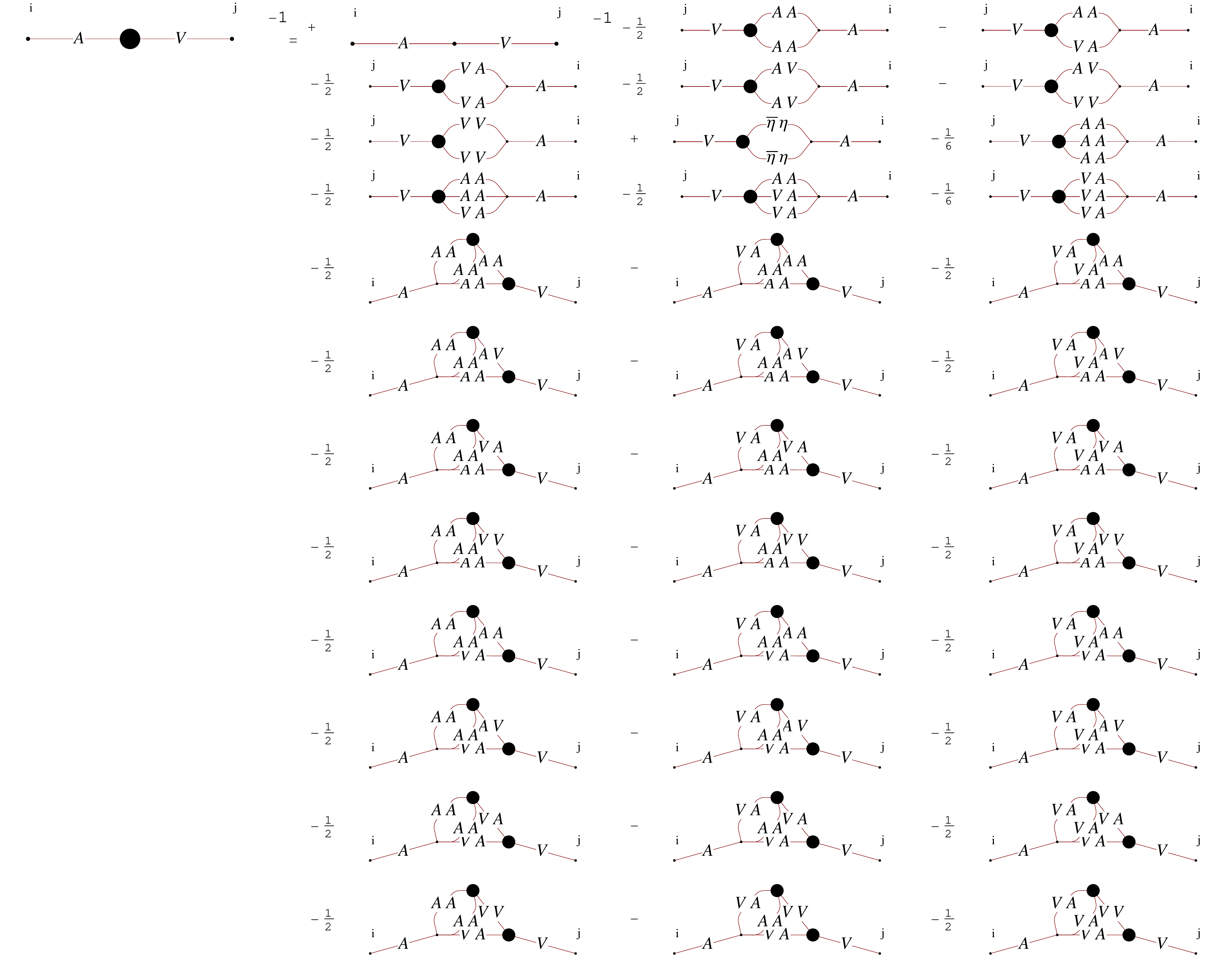}\\
\vskip10mm
\includegraphics[width=\textwidth]{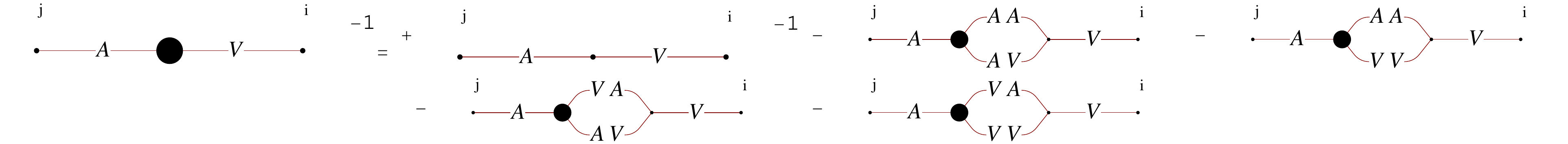}
\end{center}
\caption{\label{fig:VA-prop-DSEs}The DSEs of the $A$-$V$-mixed two-point function. The first is the $AV$-DSE and the second one the $VA$-DSE.}
\end{figure}

\begin{figure}
\begin{center}
\includegraphics[width=\textwidth]{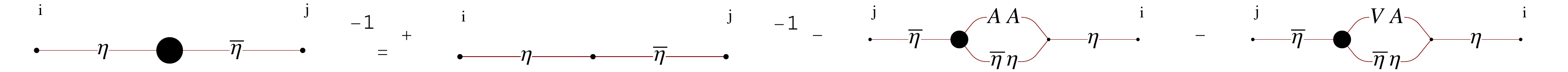}
\end{center}
\caption{\label{fig:etaeta-prop-DSEs}The DSE of the $\eta$-field two-point function.}
\end{figure}

The derivation of the DSEs for this Lagrangian is done in the standard way \cite{Alkofer:2000wg,Alkofer:2008nt}, i.~e. by variation of the path integral w.~r.~t. a field and further field differentiations. An intricacy is the mixing of the $A$- and $V$-fields, which allows already at the perturbative level additional vertices, e.~g. an $AAV$-vertex. For the $\eta$-field, on the other hand, no such vertices are allowed.
The number of diagrams in the DSEs increases drastically by the mixed propagator and the additional vertices. The task of their derivation was performed with the \textit{Mathematica} package \textit{DoDSE} \cite{Alkofer:2008nt}. The full two-point equations are given in figs. \ref{fig:AA-prop-DSEs-1L}, \ref{fig:VV-prop-DSEs-1L}, \ref{fig:VA-prop-DSEs} and \ref{fig:etaeta-prop-DSEs}.


\section{Power counting}
\label{sec:PowerCounting}

The IR part of propagators can be determined without solving the complete DSEs numerically. This technique relies on the assumption that at low momenta all dressing functions obey a power law with a specific exponent, called the infrared exponent (IRE) \cite{vonSmekal:1997vx}. For small external momentum the integrals in the propagator DSEs are dominated by small loop momenta, so that all expressions can be replaced by the corresponding IR expressions, i.~e. power laws. The integrals can be evaluated analytically or - if we are only interested in the qualitative behavior - it suffices to count the powers of all momenta in the integral, since every loop momentum will transform into an external momentum upon integration due to the lack of other mass scales. As a result the IREs of different propagators are related by so-called scaling relations, thereby reducing the number of independent IREs.

This approach was very successful in Landau gauge, where all IREs can be expressed with one parameter $\ka$ \cite{vonSmekal:1997is,vonSmekal:1997vx}. Using a bare ghost-gluon vertex its value is 
$\ka=(93 -\sqrt{1201})/98\approx 0.595$ \cite{Zwanziger:2001kw,Lerche:2002ep}. Varying dressing functions that respect the fact that the vertex does not have an
IR enhanced dressing change this value only slightly \cite{Lerche:2002ep}. Furthermore, the dependence on the kinematic configuration was also found to be mild \cite{Alkofer:2008dt}.
The first calculations took into account only the one-loop diagrams in the propagator DSEs. Later on it was shown that this IR truncation indeed is consistent
with the whole tower of DSEs \cite{Alkofer:2004it} and also RGEs \cite{Fischer:2009tn}. 
Therefore one combines the two systems of functional equations, where the decisive difference is that the DSEs feature a bare vertex in every diagram, whereas the RGEs only contain dressed quantities \cite{Fischer:2009tn,Fischer:2006vf}. The generalization to an arbitrary Lagrangian was provided in ref. \cite{Huber:2009wh}. We shortly summarize this method, since a great part of the forthcoming analysis is based on it.

Before doing so we have to 
introduce some notation. The IRE of propagators and two-point functions are given by $\de_{ij}$ and $\ka_{ij}$, respectively. This distinction will be of
importance later, because the relation between the two can be non-trivial if the Lagrangian contains mixed two-point functions, while for the case of non-mixed
two-point functions it is $\ka_{ii}=-\de_{ii}$. Higher vertex functions get the IRE $\ka_{ij\ldots}$. To distinguish different propagators and vertices we
add the corresponding fields as indices, e.~g. $\ka_{Acc}$ for the ghost-gluon vertex. Also we refer to vertices and propagators by monomials of their respective
fields, e.~g. the ghost-gluon vertex is denoted by $Acc$-vertex. If we talk about DSEs the order of the fields corresponds to the order in which the
differentiations have been performed. For instance $AVV$-DSE means the DSE of the vertex with two $V$-legs and one $A$-leg, where the first derivative was performed w.~r.~t. the
gluon field. This is important in some cases, when the order of the differentiations is
crucial. Finally, we note that we use $\ka_i:=\ka_{ii}$ for non-mixed two-point functions, e.~g. $\ka_A:=\ka_{AA}$.

In the case of non-mixing two-point functions we get the relations presented below. The calculations are done for a set of fields $\{\phi_i\}$. Considering a general diagram $v$ with $m_i$ external legs of the field $\phi_i$, the formula for its IRE $\ka_v$ is
\begin{align}\label{eq:general-IRE}
\ka_v= -\frac{1}{2}\sum_{i}m_i\delta_{{i}}+\sum_{vertices,r\geq3}n^{b}_{{i_1}\ldots {i_r}}\left(\frac{1}{2}\sum_{i}k_{{i}}^{{i_1}\ldots {i_r}}\delta_{{i}}\right)
  +\sum_{vertices,r\geq3}n^{d}_{{i_1}\ldots  {i_r}}\left(\ka_{{i_1}\ldots {i_r}}+\frac{1}{2}\sum_{i}k_{{i}}^{{i_1}\ldots {i_r}}\delta_{{i}}\right),
\end{align}
where $\de_i$ and $\ka_{{i_1}\ldots {i_r}}$ are the propagator and vertex IREs, $n^{b/d}_{{i_1}\ldots {i_r}}$ is the number of bare/dressed $\phi_{i_1}\ldots\phi_{i_r}$-vertices and $k_{{i}}^{{i_1}\ldots {i_r}}$ gives the number of $\phi_i$-legs of a $\phi_{i_1}\ldots\phi_{i_r}$-vertex. Furthermore it was found that DSEs and RGEs yield bounds on the last two terms, namely
\begin{align} \label{eq:props-vert-ineq}
\ka_{{i_1}\ldots {i_r}}+\frac{1}{2}\sum_{i}k_{{i}}^{{i_1}\ldots {i_r}}\delta_{{i}} \geq 0,
\end{align}
which exists for every possible vertex with the IRE $\ka_{{i_1}\ldots {i_r}}$, and
\begin{align} \label{eq:props-ineq}
\frac1{2} \sum_{i} k_{i}^{{i_1} \ldots {i_r}} \de_{i} \geq 0,
\end{align}
which only
relates to the vertices appearing in the Lagrangian.
Details on these two equations are given in Appendix \ref{subsec:non-neg}.
Thus the last two terms in \eref{eq:general-IRE} can be dropped to get the lowest possible IRE of the diagram $v$:
\begin{align}
\delta_{v,max}= -\frac{1}{2}\sum_{i}m_{i}\delta_{{i}}.
\end{align}
It depends only on the number of external legs and is therefore the same for all diagrams of a DSE.
Using this expression in the analysis of leading diagrams in the propagator DSEs, one arrives at the following inequality:
\begin{align}\label{eq:leadingDiagram}
-\frac1{2}\hat{n}^b_{{i_1} \ldots {i_r}} \sum_{j} \de_{j} k_{j}^{{i_1} \ldots {i_r}} \geq 0,
\end{align}
where the hat indicates that this equation is only valid for the leading diagram(s). In every diagram of a DSE is exactly one bare vertex, so that there is only one $n^b=1$, while the others are zero.
Since this inequality is just the opposite of \eref{eq:props-ineq}, it has to be saturated, i.~e.
\begin{align}\label{eq:scalingRelation}
\frac1{2} \hat{n}^b_{{i_1} \ldots {i_r}} \sum_{j}\de_{j} k_{j}^{{i_1} \ldots {i_r}}=0.
\end{align}
An immediate consequence is
\begin{align}
\hat{\ka}_{{i_1} \ldots {i_r}}=0,
\end{align}
i.~e. the vertex corresponding to $\hat{n}^b_{{i_1} \ldots {i_r}}$ does not scale in the IR. This is a direct consequence of the Fischer-Pawlowski consistency condition \cite{Fischer:2006vf}, namely that there is one bare vertex in the DSEs and none in the RGEs, but both equations should yield the same result. In this analysis it was not specified which diagram was assumed to be leading. This can be done now, i.~e. one tries all possible realizations of \eref{eq:scalingRelation}. Since there are only as many possibilities as bare vertices in the Lagrangian this is a manageable task. In other words, one can read off possible scaling relations directly from the Lagrangian and take only those that yield non-trivial results \cite{Huber:2009wh}. Let us illustrate this approach in standard Landau gauge: We only have three bare vertices: the ghost-gluon vertex, the three-gluon vertex and the four-gluon vertex. The two gluonic vertices yield the same inequality, which even simplifies our task. The two possibilities that remain are:
\begin{align}
\de_A&=0 \quad \quad \text{(from gluonic vertices)},\\
&\text{or}\nnnl
\de_A+2\de_c&=0 \quad \quad \text{(from ghost-gluon vertex)}.
\end{align}
The first possibility corresponds to the perturbative (trivial) solution, i.~e. all Green functions are bare, whereas the second solution yields the known scaling relation.
As this simple example shows, the method presented in ref. \cite{Huber:2009wh} provides an easy method to derive a scaling relation. Its use is especially convenient for more complicated Lagrangians like Yang-Mills theory in the maximally Abelian gauge \cite{Huber:2009wh}.

The task is now to generalize these results to a Lagrangian with mixed propagators.
An important consequence of this mixing is that a propagator is not the inverse of the two-point function, but there exists a matrix relation between propagators $D^{\phi\phi}$ and two-point functions $\Gamma^{\phi\phi}$,
\begin{align}
D^{\phi\phi}=(\Gamma^{\phi\phi})^{-1},
\end{align}
where $\phi$ can be any field.
This is the reason why we assign different IREs to propagators and two-point functions, namely $\de_{ij}$ and $\ka_{ij}$, respectively, and lies at the heart of the complication of the method.

Whereas eqs. (\ref{eq:props-vert-ineq}) and (\ref{eq:props-ineq}) remain valid, as exemplified in Appendix \ref{sec:general-IRE}, \eref{eq:general-IRE} undergoes some modifications, since the topological relations used in its derivation have changed; see eqs. (\ref{eq:numberOfLoops}) and (\ref{eq:relation-verts-props}). The details of the calculation are given in App. \ref{sec:general-IRE}. From now on we work with general dimension $d$. As a result we obtain for the lower bound on the IRE of a diagram
\begin{align}\label{eq:lowerBoundMod}
\ka_{v,max} = & \left(\frac{d}{2}-2\right)\left(1-\frac{1}{2}\sum_{i}m_{i}\right)-\frac{1}{2}\sum_{i}m_{i}\delta_{{i}}+\frac1{2}n_{AV}(2\de_{AV}-\de_A-\de_V).
\end{align}
The new term is the last one, by which the formula depends on the number of internal mixed propagators. Also \eref{eq:leadingDiagram} gets modified:
\begin{align}\label{eq:leadingDiagramMod}
\ka_i+\frac1{2}\sum_j \de_j m_j-n^b_{{i_1}\ldots {i_r}}\left(\left(\frac{d}{4}-1\right)(r-2)+\frac{1}{2}\sum_{j}k_{{j}}^{{i_1}\ldots {i_r}}\delta_{{i}}\right)  -\frac1{2}(2\de_{AV}-\de_A-\de_V)\left( n_{AV}+\sum_{\substack{dressed\\vertices}} n^d_{{i_1} \ldots {i_r}} \bar{k}^{{i_1} \ldots {i_r}}_{AV} \right) \geq0,
\end{align}
where $\bar{k}^{{i_1} \ldots {i_r}}_{AV}$ indicates the number of times a mixed propagator is contained in the diagram that determines the IRE of the vertex $\phi_{i_1} \ldots \phi_{i_r}$.

An issue we have to address is the role of the bare two-point functions appearing in the two-point DSEs. In the following analysis we ignore them. For the gluon propagator this makes no difference at all in the analysis, while for the other propagators this is of vital importance. The reason is that including the bare two-point function would restrict the IRE of the two-point function to negative values. However, for the Faddeev-Popov ghost propagator, the fermionic auxiliary field and the real part of the bosonic auxiliary field it is well justified to neglect it in an IR analysis due to the horizon condition, which can be used to show that the ghost two-point function has to vanish at zero momentum and thus the bare two-point function has to be canceled \cite{Zwanziger:1992qr,Zwanziger:1993dh,Zwanziger:2009je}. 
This is the way the horizon condition enters in our calculations. A similar mechanism for the mixed propagator is not known. We comment on this and the cancelation of the imaginary part of the bosonic auxiliary field $V$ at the ends of subsec. \ref{ssec:CaseII} and \ref{ssec:CaseIII}.

\section{Propagators and vertex functions in the infrared}
\label{sec:Propagators}

Normally one does not have to distinguish between different tensors for power counting, since automatically the most divergent tensor is taken into account. However, when the relation between the propagators and two-point functions is not trivial as here, more care has to be taken.
To keep the number of tensors manageable we use the following truncation of the tensor basis: We only take into account those tensors of two-point functions that appear in the Lagrangian. 
We will see that already in this minimal truncation a non-trivial $VV$-propagator arises, in the sense that there are two independent parts with different dressing functions. Thus they can behave differently and both of them have to be taken into account.

Explicitly the matrix of the mixing two-point functions is given by
\begin{align}
\Gamma^{\phi\phi}&=\begin{pmatrix}
\Gamma^{AA} & \Gamma^{AV}\\
\Gamma^{VA} & \Gamma^{VV}
\end{pmatrix},\\
\Gamma^{AA,ac}_{\mu\nu}&=\delta^{ac}p^2 c_A^\bot(p^2) P_{\mu\nu}+\delta^{ac}\frac{1}{\xi} c_A^\parallel(p^2) p_\mu p_\nu,\\
\Gamma^{VV,abcd}_{\mu\nu}&=\delta^{ac}\delta^{bd}p^2 c_V(p^2) g_{\mu\nu},\\
\Gamma^{AV,cab}_{\mu\nu}&=f^{cab} i\, p^2 c_{AV}(p^2) g_{\mu\nu},
\end{align}
and that of the propagators by ($\xi=0$)
\begin{align}
D^{\phi\phi}&=
\begin{pmatrix}
D^{AA} & D^{AV}\\
D^{V A} & D^{VV}
\end{pmatrix},\\
D^{AA,ab}_{\mu\nu}&=\delta^{ab}\frac{1}{p^2} P_{\mu\nu} \frac{c_V(p^2)}{c_A^\bot(p^2) c_V(p^2)+2 N\, c^2_{AV}(p^2)},\\
D^{VV,abcd}_{\mu\nu}&=\frac{1}{p^2}\frac{1}{c_V(p^2)}\delta^{ac}\delta^{bd}g_{\mu\nu}-f^{abe}f^{cde}\frac1{p^2}P_{\mu\nu}\frac{2  c_{AV}^2(p^2)}{c_A^\bot(p^2) c_V^2(p^2)+2N\, c_{AV}^2(p^2) c_V(p^2)},\\
D^{AV,abc}&=-i\,f^{abc}\frac1{p^2}P_{\mu\nu}\frac{\sqrt{2} c_{AV}(p^2)}{c_A^\bot(p^2) c_V(p^2)+2N\,c_{AV}^2(p^2)}.
\end{align}
For the fermionic ghost the standard relation is valid:
\begin{align}
D^{\eta\bar{\eta},ab}_{cd}=(\Gamma^{\eta\bar{\eta},ab}_{cd})^{-1}=-\de^{ab}\de_{cd}\frac{c_{\eta}(p^2)}{p^2}.
\end{align}
The functions $c_{ij}$ are the dressing functions of the two-point functions with the IREs $\ka_{ij}$ defined as
\begin{align}
c_{ij}(p^2)\overset{IR}{=}d_{ij}\,(p^2)^{\ka_{ij}}.
\end{align}
Similar to the IREs we define $d_{ii}=d_i$ and $c_{ii}=c_i$.
Because of the matrix inversion the determinant of the two-point function matrix appears, $det\,C=c_A^\bot(p^2) c_V(p^2)+2N\,c_{AV}^2(p^2)$, which forces us to distinguish between several cases:
\begin{enumerate}
 \item[I:] $c_{AV}^2>c_A c_V \leftrightarrow \ka_A+\ka_V>2\ka_{AV}$
 \item[II:] $c_A c_V>c_{AV}^2 \leftrightarrow 2\ka_{AV}>\ka_A+\ka_V$
 \item[III:] $c_{AV}^2\sim c_A c_V \leftrightarrow \ka_A+\ka_V=2\ka_{AV}$, no cancelations
 \item[IV:] $c_{AV}^2\sim c_A c_V \leftrightarrow \ka_A+\ka_V=2\ka_{AV}$, cancelations
\end{enumerate}
If the IREs of the two expressions in the determinant are the same, it is possible that the IR leading parts cancel. Thus we have to take into account this case explicitly as case IV.

We will now determine the form of the propagators for each of the four cases.

\subsection{Case I: $\ka_A+\ka_V>2\ka_{AV}$}

By assumption the dressing function $c_{AV}$ dominates in the numerator. Therefore the propagators take the following form:
\begin{align}
D^{AA,ab}_{\mu\nu}&=\delta^{ab}\frac{1}{p^2} P_{\mu\nu} \frac{c_V(p^2)}{2 N\, c^2_{AV}(p^2)},\\
D^{VV,abcd}_{\mu\nu}&=\frac{1}{p^2}\frac{1}{c_V(p^2)}\left(\delta^{ac}\delta^{bd}g_{\mu\nu}-f^{abe}f^{cde}P_{\mu\nu}\frac{1}{N}\right),\\
D^{AV,abc}&=-i\,f^{abc}\frac1{p^2}P_{\mu\nu}\frac{1}{\sqrt{2}N\,c_{AV}(p^2)}.
\end{align}
Note that both tensors contribute to the $VV$-propagator, since they have the same IRE.
The relations between the IREs of propagators and two-point functions are
\begin{align}
\de_A&=\ka_V-2\ka_{AV},\\
\de_V&=-\ka_V,\\
\de_{AV}&=-\ka_{AV}.
\end{align}
With these expressions we can determine the value of the additional term that appears in eqs. (\ref{eq:lowerBoundMod}) and (\ref{eq:leadingDiagramMod}):
\begin{align}
2\de_{AV}-\de_A-\de_V=-2\ka_{AV}-\ka_V+2\ka_{AV}+\ka_V=0.
\end{align}
Thus the two equations stay the same as in the case with no mixing propagators. The lower bound is then as usual,
\begin{align}
\ka_{v,max} = & \left(\frac{d}{2}-2\right)\left(1-\frac{1}{2}\sum_{i}m_{i}\right)-\frac{1}{2}\sum_{i}m_{i}\delta_{{i}},
\end{align}
and \eref{eq:leadingDiagramMod} takes the form
\begin{align}\label{eq:leadingDiagramModI}
\ka_i+\frac1{2}\sum_j \de_j m_j-n^b_{{i_1}\ldots {i_r}}\left(\left(\frac{d}{4}-1\right)(r-2)+\frac{1}{2}\sum_{j}k_{{j}}^{{i_1}\ldots {i_r}}\delta_{{i}}\right)\geq0.
\end{align}

\begin{figure}
\begin{minipage}[c]{0.45\linewidth}
\includegraphics[width=\textwidth]{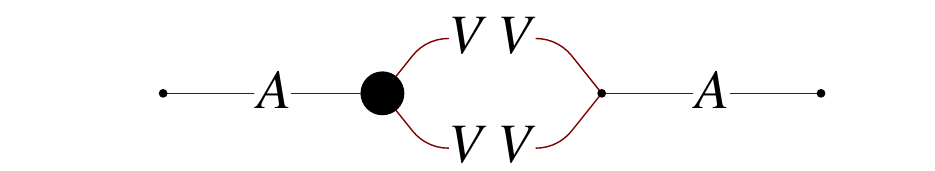}
\end{minipage}
\begin{minipage}[c]{0.45\linewidth}
\includegraphics[width=\textwidth]{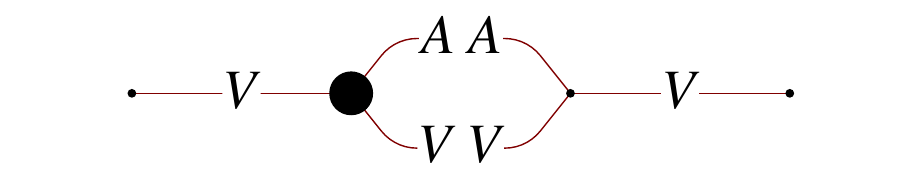}
\end{minipage}
\caption{\label{fig:AA-WW-loops}Diagrams of the $AA$- and $VV$-DSEs used explicitly in the text.}
\end{figure}

It is now a proof of only a few lines to show that case I corresponds to an inconsistent assumption. In the $VA$-DSE, see \fref{fig:VA-prop-DSEs}, we only have bare $AVV$-vertices. Thus \eref{eq:leadingDiagramModI} becomes
\begin{align}
\ka_{AV}+\frac1{2}(\de_A+\de_V)-\frac1{2}(\de_A+2\de_V)-\frac1{2}\left(\dhalf-2\right) \geq0 \quad \Rightarrow \quad
2\ka_{AV} \geq-\ka_V+ \dhalf-2.
\end{align}
From the $VV$-loop of the $AA$-DSE, depicted in \fref{fig:AA-WW-loops}, we get
\begin{align}
\ka_A&\leq-2\ka_V+\frac{d}{2}-2 \quad \Rightarrow \quad \ka_A+\ka_V\leq-\ka_V+\dhalf-2,
\end{align}
where $\ka_{AVV}\leq0$ from the $AVV$-DSE was used.
Using these expressions in the defining assumption of case I, we find an unsatisfiable inequality:
\begin{align}
-\ka_V+\dhalf-2\geq\ka_A+\ka_V&>2\ka_{AV}\geq-\ka_V+\dhalf-2,\nnnl
0&>0.
\end{align}
Thus case I can be disregarded
except one potential caveat.
For case I we cannot exclude that the mixed two-point function might not be determined 
by the loop integrals. Therefore we also discuss the case in which the tree-level term is the IR dominant part, which corresponds to $\ka_{AV}=-1$, since the tree-level expression is constant. In the $VV$-DSE there is only one bare vertex, the $AVV$-vertex. Thus \eref{eq:leadingDiagramModI} yields
\begin{align}
\de_A+2\de_V+\dhalf-2\leq 0.
\end{align}
The $VV$-DSE itself leads from the $AA$-$VV$-loop, as given in \fref{fig:AA-WW-loops}, to
\begin{align}
\de_A+2\de_V+\dhalf -2\geq 0.
\end{align}
Together these two inequalities give
\begin{align}
\de_A+2\de_V+\dhalf-2=0.
\end{align}
Inserting the expressions for the propagator IRE, we arrive at
\begin{align}
\ka_V=\dhalf-2-2\ka_{AV}=\dhalf.
\end{align}
The last inequality we need we get from the $AA$-DSE, namely from the $VV$-loop as depicted in \fref{fig:AA-WW-loops}:
\begin{align}
\ka_A\leq2\de_V+\dhalf-2=-2\ka_V+\dhalf-2.
\end{align}
Now we insert these equations into the defining relation of case I:
\begin{align}
-\ka_V+\dhalf-2\geq\ka_A+\ka_V>&2\ka_{AV}\nnnl
-2>&-2.
\end{align}
We see that also the case
of a constant two-point function leads to an inconsistency.

\subsection{Case II: $\ka_A+\ka_V<2\ka_{AV}$}
\label{ssec:CaseII}

Here the dressing function combination $c_A c_V$ dominates in the numerator:
\begin{align}
D^{AA,ab}_{\mu\nu}&=\delta^{ab}\frac{1}{p^2} P_{\mu\nu} \frac{1}{c_A^\bot(p^2)},\\
D^{VV,abcd}_{\mu\nu}&=\frac{1}{p^2}\frac{1}{c_V(p^2)}\left(\delta^{ac}\delta^{bd}g_{\mu\nu}-f^{abe}f^{cde}P_{\mu\nu}\frac{2  c_{AV}^2(p^2)}{c_A^\bot(p^2) c_V(p^2)}\right)
\rightarrow \frac{1}{p^2}\frac{1}{c_V(p^2)}\delta^{ac}\delta^{bd}g_{\mu\nu},\\
D^{AV,abc}&=-i\,f^{abc}\frac1{p^2}P_{\mu\nu}\frac{\sqrt{2} c_{AV}(p^2)}{c_A^\bot(p^2) c_V(p^2)}.
\end{align}
Only the first tensor of the $VV$-propagator contributes at leading order, since the second one is additionally suppressed as one can see from the defining
condition $\ka_A+\ka_V<2\ka_{AV}$ of 
the here considered case II. The relations between the IREs are
\begin{align}\label{eq:delta-kappa-II}
\de_A&=-\ka_A,\\
\de_V&=-\ka_V,\\
\de_{AV}&=\ka_{AV}-\ka_A-\ka_V.
\end{align}
With this the new term in the formula for the lower bound on an IRE becomes
\begin{align}
2\de_{AV}-\de_A-\de_V=2\ka_{AV}-2\ka_A-2\ka_V+\ka_{A}+\ka_V=2\ka_{AV}-\ka_A-\ka_V>0,
\end{align}
where we used the defining assumption of case II for the last step. Thus all diagrams that feature mixed propagators are additionally suppressed. This allows one to determine the value $\ka:=-\ka_V=-\ka_\eta$ immediately, since if we disregard in the IR all diagrams that contain an $AV$-propagator or vertices that necessarily contain an $AV$-propagator like the $AAV$-vertex, the usual Landau gauge system of DSEs remains. Taking into account that here the $VV$-propagator has exactly the same form as the propagator of the fermionic ghost, the numerical factors from color and Lorentz indices add up in such a way that effectively only contributions from the Faddeev-Popov ghost contribute, as expected. Remember that we could not integrate out the $V$-ghost as the other ghosts, because it coupled to the gluon. In case II, however, this coupling is IR suppressed and does not play a role at leading IR order. The auxiliary field contributions cancel each other and the value of $\ka$ is calculated as $0.595353$ as for the standard Faddeev-Popov action. Since at leading IR order the DSEs are the same as in normal Landau gauge, we can use other results from it as well, e.~g. the formula for the IRE of a vertex with $2n$ ghost- and $m$ gluon-legs \cite{Alkofer:2004it,Huber:2007kc}:
\begin{align}
\ka_{2n,m}=(n-m)\ka+(1-n)\left(\mhalf{d}-2\right).
\end{align}

\begin{figure}
\begin{center}
\includegraphics[width=0.45\textwidth]{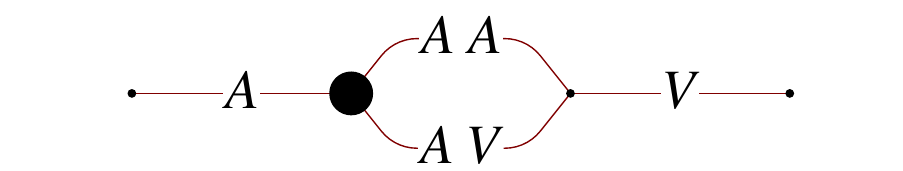}
\includegraphics[width=0.45\textwidth]{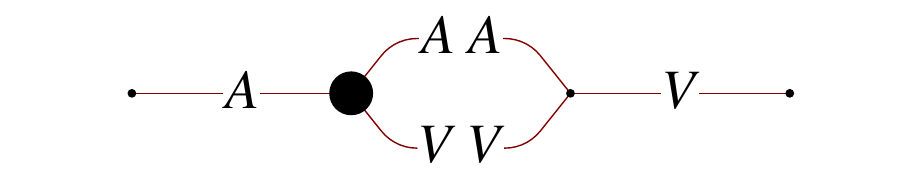}\\
\vskip10mm
\includegraphics[width=0.45\textwidth]{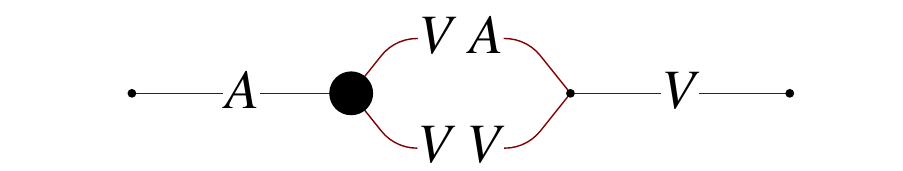}
\end{center}
\caption{\label{fig:VA-loops}Diagrams of the $VA$-DSE used explicitly in the text.}
\end{figure}

The IRE of the mixed two-point function can be calculated from its DSE. Here the importance of diagrams in the IR is determined by the number of $AV$-propagators, where one has to keep in mind that vertices exist that must have at least one mixed propagator in their DSEs. The one-loop diagrams of the $VA$-DSE, as depicted in \fref{fig:VA-loops}, have the following IREs:
\begin{align}\label{eq:VA-diagrams1}
\de_A+\de_{AV}+\ka_{AAA},
 \quad \de_A+\de_V+\ka_{AAV}, 
 \quad \de_V+\de_{AV}+\ka_{AVV}.
\end{align}
The fourth diagram existing in the $VA$-DSE is subleading, since it has two mixed propagators.
$\ka_{AVV}$ is zero due to the scaling relation and the IREs of $\ka_{AAA}$ and $\ka_{AAV}$ can be determined from their DSEs as
\begin{align}
\ka_{AAA}&=3\de_V,\\
\ka_{AAV}&=\de_{AV}+2\de_V.
\end{align}
Plugging this into \eref{eq:VA-diagrams1} and using \eref{eq:delta-kappa-II} as well as the scaling relation $\ka_A+2\ka_V=0$, we get $\ka_{AV}$ for all three diagrams, which proves that they are at leading IR order.
For the calculation of $\ka_{AV}$ we have to deal with several unknowns: The coefficients of the dressing functions ($d_{AV}$, $d_{V}$, $d_A$) and the IREs
$\ka_V$ and $\ka_{AV}$. However, $d_{AV}$ always drops out of the equations and $d_V$ and $d_A$ combine to $d_A d_V^2$, which can be calculated from the other
propagator DSEs as $0.0267784$. That leaves only $\ka_{AV}$. For a practical calculation we further truncate the system, since we cannot take into account all
leading diagrams. We only take the $AV$-$VV$-loop into account, since it only contains $AVV$-vertices. The IRE of the $AVV$-vertex is $0$, i.~e. the bare diagram
contributes to leading order. There are in principal other diagrams with the same order, but similar to the Faddeev-Popov ghost-gluon vertex, they 
can be neglected. They would correspond to a higher-loop correction. For the same reason we 
can neglect other diagrams in the $VA$-DSE, since their IR leading contributions come from one-loop diagrams with IR enhanced vertices and plugging the IR leading parts of the vertices into the two-point DSE yields two-loop diagrams. In other words we take only into account the order $d_A d_V^2$ and neglect everything of order $(d_A d_V^2)^2$ and higher. However, one should keep in mind that such truncations are only needed to calculate a numeric value for the IREs and that the qualitative features and the scaling relations themselves are derived for the full system without truncations.

\begin{figure}
\includegraphics{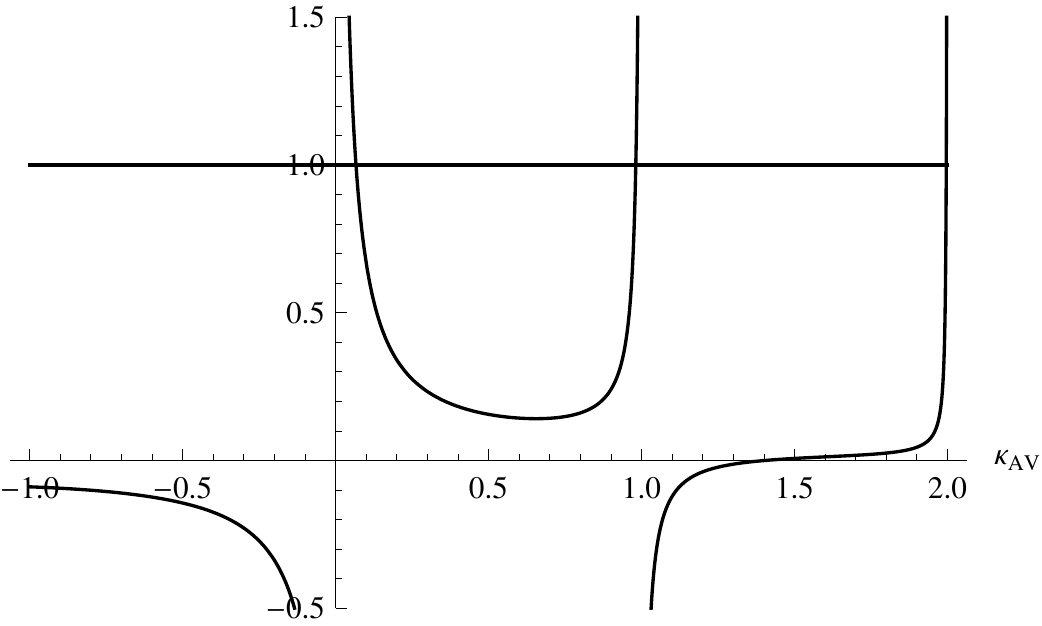}
\caption{\label{fig:kappaAV}The plot to determine solutions for $\ka_{AV}$. On the 
$Y$-axis we plot the left- and right-hand sides of \eref{eq:kappaAV}.}
\end{figure}

The equation to determine $\ka_{AV}$ is obtained by projecting transversely in Lorentz space and by $f^{abc}$ in color space:
\begin{align}
(d-1)(p^2)^{\ka_{AV}+1} d_{AV}=- g^2\,N\int \frac{d^dq}{(2\pi)^d} \frac{d_{AV}  ((p+q)^2)^{-1+\de_V} (q^2)^{-1+\de_{AV}} \left(p^2 q^2-(p\,q)^2\right)}{\sqrt{2} p^2 d_A d_V^2},
\end{align}
The coefficient $d_{AV}$ drops out of the equation. The final equation that has to be solved numerically is
\begin{align}\label{eq:kappaAV}
1&=\frac{0.0513093 \Gamma(1-\ka_{AV}) \Gamma(0.595353+\ka_{AV})}{\Gamma(1.40465-\ka_{AV}) \Gamma(1+\ka_{AV})}-\frac{0.0370825 \Gamma(-\ka_{AV}) \Gamma(0.595353+\ka_{AV})}{\Gamma(1.40465-\ka_{AV}) \Gamma(2+\ka_{AV})}-\nnnl
&-\frac{0.0513093 \Gamma(-\ka_{AV}) \Gamma(1.59535+\ka_{AV})}{\Gamma(0.404647-\ka_{AV}) \Gamma(2+\ka_{AV})}-\frac{0.0210772 \Gamma(-1-\ka_{AV}) \Gamma(0.595353+\ka_{AV})}{\Gamma(1.40465-\ka_{AV}) \Gamma(3+\ka_{AV})}-\nnnl
&-\frac{0.0247217 \Gamma(-1-\ka_{AV}) \Gamma(1.59535+\ka_{AV})}{\Gamma(0.404647-\ka_{AV}) \Gamma(3+\ka_{AV})}-\frac{0.0513093 \Gamma(-1-\ka_{AV}) \Gamma(2.59535+\ka_{AV})}{\Gamma(-0.595353-\ka_{AV}) \Gamma(3+\ka_{AV})}-\nnnl
&-\frac{0.0237307 \Gamma(-2-\ka_{AV}) \Gamma(0.595353+\ka_{AV})}{\Gamma(1.40465-\ka_{AV}) \Gamma(4+\ka_{AV})}-\frac{0.0210772 \Gamma(-2-\ka_{AV}) \Gamma(1.59535+\ka_{AV})}{\Gamma(0.404647-\ka_{AV}) \Gamma(4+\ka_{AV})}-\nnnl
&-\frac{0.0370825 \Gamma(-2-\ka_{AV}) \Gamma(2.59535+\ka_{AV})}{\Gamma(-0.595353-\ka_{AV}) \Gamma(4+\ka_{AV})}+\frac{0.0513093 \Gamma(-2-\ka_{AV}) \Gamma(3.59535+\ka_{AV})}{\Gamma(-1.59535-\ka_{AV}) \Gamma(4+\ka_{AV})}.
\end{align}
As one can see from \fref{fig:kappaAV} there are several solutions. A constraint for $\ka_{AV}$ is $2\ka_{AV}>\ka_V+\ka_A=-\ka_V+d/2-2$.
The smallest solutions fulfilling this constraint are $0.0668776$ and $0.981386$. Further solutions always just a little bit below integer numbers follow. At this point we cannot say which one is correct. In our truncation all of them are admissible.

An interesting observation is the fact that the $AV$-$VV$-loop (see \fref{fig:VA-loops}) is automatically at leading order:
\begin{align}
\de_V+\de_{AV}+\ka_{AVV}=-\ka_V+\ka_{AV}-\ka_A-\ka_V=\ka_{AV}.
\end{align}
Therefore the tree-level term has to be canceled in order to allow a solution with $\ka<2$. For the $VV$-two-point function we can employ the horizon condition in the same way as for the Faddeev-Popov ghost to explain the required cancelation of the tree-level term in the DSE, since their DSEs are the same in the IR.

\subsection{Case III: $\ka_A+\ka_V=2\ka_{AV}$, no cancelations}
\label{ssec:CaseIII}

Writing $det\, C=  c_A c_V +  2 N c_{AV}^2= (p^2)^{2\ka_{AV}} det\,D= (p^2)^{\ka_A+\ka_V} det\,D$,  we get 
\begin{align}\label{eq:propsIII}
D^{AA,ab}_{\mu\nu}&=\delta^{ab}\frac{1}{p^2} P_{\mu\nu} \frac{d_V}{(p^2)^{\ka_A}}\frac{1}{det\,D},\\
D^{VV,abcd}_{\mu\nu}&=\frac{1}{p^2}\frac{1}{c_V(p^2)}\left(\delta^{ac}\delta^{bd}g_{\mu\nu}-f^{abe}f^{cde}P_{\mu\nu}\frac{2  d_{AV}^2}{det\,D}\right),\\
D^{AV,abc}&=-i\,f^{abc}\frac1{p^2}P_{\mu\nu}\frac{d_{AV}}{(p^2)^{\ka_{AV}}}\frac{\sqrt{2}}{det\,D}.
\end{align}
If no cancelations occur $det \, D$ is just a momentum independent constant and it assumes a similar role as the coefficients of the power laws, i.~e. it is an unknown constant that has to be determined. The relations between the IREs are
\begin{align}
\de_A=-\ka_A,\\
\de_V=-\ka_V,\\
\de_{AV}=-\ka_{AV}.
\end{align}
The additional term in the formulas for the IREs, eqs. (\ref{eq:lowerBoundMod}) and (\ref{eq:leadingDiagramMod}), becomes
\begin{align}
2\de_{AV}-\de_A-\de_V=-2\ka_{AV}+\ka_{A}+\ka_V=0,
\end{align}
where we used the defining assumption of case III. Thus again the new terms vanish, i.~e. we can directly get the scaling relation from the vertices in the Lagrangian, if we assume that the tree-level two-point function in the $VV$-DSE gets canceled in a similar way as that for the other IR enhanced fields. The only non-trivial solution arises from the $A\bar{\eta}\eta$- and $AVV$-vertices:
\begin{align}
\ka:=\ka_V=\ka_\eta, \quad \ka_A+2\ka=\dhalf-2,\quad \ka_{AVV}=\ka_{A\eta\eta}=0.
\end{align}
The IRE of the mixed two-point function can be calculated from the defining assumption as
\begin{align}
\ka_{AV}=-\frac{\ka}{2}+\frac{d}{4}-1.
\end{align}
In all DSEs the diagrams with a bare $A\bar{\eta}\eta$- or $AVV$-vertex are leading and the formula for the IRE of a vertex is
\begin{align}
\ka_{2n,m}=(n-m)\ka+(1-n)\left(\mhalf{d}-2\right),
\end{align}
where $m$ is the number gluon legs and $n$ the number of legs of ghosts or auxiliary fields.

Thus the qualitative behavior is the same as case II, i.~e. the propagators of the Faddeev-Popov ghost and the auxiliary fields are IR enhanced and the gluon propagator is IR suppressed, as is the mixed propagator. The difference between case II and III is that the mixed propagator is more IR suppressed with respect to $\ka$ in case II, since there it holds that $\de_{AV}>\ka/2$, whereas in case III it is $\ka/2$.

Having found a scaling relation does, however, not yet mean that we can be sure there really exists a corresponding scaling solution. A first indication of its existence is a solution for the IRE $\ka$. This is an intricate task as the system of equalities that has to be solved involves the coefficients of the power laws, the $d_{ij}$, in a non-linear way. In case II only the IRE $\ka$ appeared non-linearly and consequently one can obtain all solutions within a given interval. In the present case we were able to find several possible solutions with $0<\ka<1$, but all of them are in contradiction to requirements such as the positivity of $d_c$ and $d_A$. Thus we cannot guarantee the existence of a solution in case III.

For case III it is necessary that the tree-level term in the $AV$-DSE gets canceled by some mechanism that may be related to the horizon condition. If this is not the case the $AV$-$VV$-loop can only be leading for $\ka\geq2$ or - if the tree-level term leads - we get $\ka=2$.

\subsection{Case IV: $\ka_A+\ka_V=2\ka_{AV}$, cancelations}

In the case that the IREs fulfill the relation $\ka_A+\ka_V=2\ka_{AV}$ it is possible that the leading terms in the determinant $det\,C=c_A c_V +  2 N c_{AV}^2= (p^2)^{2\ka_{AV}} det\,D= (p^2)^{\ka_A+\ka_V} det\,D$ cancel, i.~e. $d_A d_V=-2\,N\,d_{AV}^2$. 
As the IRE of the next leading term is undetermined yet we introduce  $det\,D=d_D (p^2)^{\ka_D}$, with $\ka_D>0$. This bound comes from having factored out the IREs of the leading dressing functions in the determinant: The remaining pieces are constant or IR suppressed. In case IV the constant pieces cancel each other and something IR suppressed remains:
\begin{align}
 det\,C=c_A c_V + 2 N c_{AV}^2=(p^2)^{\ka_A+\ka_V}(d_A d_V+2N d_{AV}^2 + d_D (p^2)^{\ka_D}+\ldots )=(p^2)^{\ka_A+\ka_V}( d_D (p^2)^{\ka_D}+\ldots).
\end{align}

  The propagators are then almost the same as in case III, except for the $VV$-propagator, where now the first term is suppressed:
\begin{align}
D^{AA,ab}_{\mu\nu}&=\delta^{ab}\frac{1}{p^2} P_{\mu\nu} \frac{d_V}{(p^2)^{\ka_A}}\frac{1}{det\,D},\\
D^{VV,abcd}_{\mu\nu}&=\frac{1}{p^2}\frac{1}{c_V(p^2)}\left(\delta^{ac}\delta^{bd}g_{\mu\nu}-f^{abe}f^{cde}P_{\mu\nu}\frac{2  d_{AV}^2}{det\,D}\right)
\rightarrow \frac{1}{p^2}\frac{1}{c_V(p^2)}\left(-f^{abe}f^{cde}P_{\mu\nu}\frac{2  d_{AV}^2}{det\,D}\right),\\
D^{AV,abc}&=-i\,f^{abc}\frac1{p^2}P_{\mu\nu}\frac{d_{AV}}{(p^2)^{\ka_{AV}}}\frac{\sqrt{2}}{det\,D}.
\end{align}
The relations between the IREs of propagators and two-point functions are in this case
\begin{align}
\de_A=-\ka_A-\ka_D,\\
\de_V=-\ka_V-\ka_D,\\
\de_{AV}=-\ka_{AV}-\ka_D.
\end{align}
Thus we have again $2\de_{AV}-\de_A-\de_V=0$.

The inequalities relating the different IREs of propagators, \eref{eq:props-ineq}, and vertices, \eref{eq:props-vert-ineq}, remain valid, since there does not appear any IRE of two-point functions. However, the appearance of $\ka_D$ influences the remaining power counting as follows.
From \eref{eq:leadingDiagramMod}, which gives a bound from the analysis of propagator equations, we get for the $VV$-propagator:
\begin{align}
\ka_V+\de_V-\frac1{2}\left( \de_A+2\de_V\right)\geq&0,\nnnl
\ka_A+2\ka_V+\ka_D\geq&0.
\end{align}
We chose the $VV$-propagator because there is only one possible bare vertex in its DSE, contrary to the $AA$-DSE.
On the other hand, from counting the $VV$-$AA$-loop in the $VV$-DSE, see \fref{fig:AA-WW-loops}, we get another bound:
\begin{align}
\ka_V\leq&\de_V+\de_A+\ka_{AVV}\, ,\nnnl
\ka_A+2\ka_V+2\ka_D\leq&\ka_{AVV}\leq0,
\end{align}
where we used $\ka_{AVV}\leq0$ from the DSE of the $AVV$-vertex.
Without $\ka_D$ we could now infer the usual scaling relation. However, combining these two inequalities yields now some additional information on $\ka_D$:
\begin{align}
-\ka_D\leq\ka_A+2\ka_V&\leq-2\ka_D,\nnnl
\ka_D\geq2\ka_D.
\end{align}
This inequality cannot be fulfilled, since $\ka_D$ is non-negative and we are back at $\ka_D=0$ (case III). Thus case IV does not yield a solution.

\section{Conclusions}
\label{sec:conclusions}

The analysis of the IR regime of the Gribov-Zwanziger Lagrangian as performed in this article requires a method to deal with the mixing of the fields at the
two-point level. Some general formulas necessary for this, like the lower bound on the IREs of vertex functions, were given in Sec. \ref{sec:PowerCounting}.
However, the complete analysis is 
severely complicated by the mixing of the fields, which leads to a matrix relation between two-point functions and propagators. This necessitates the distinction between four different cases, which assume different combinations of dressing functions to be more important than others. Only two of these assumptions allowed a consistent solution: We could explicitly exclude that the dressing functions of the mixed two-point functions dominate in the IR and thereby lead to a completely new IR behavior of the gluon and Faddeev-Popov ghost. The second dismissable case allowed for cancelations in the determinant of two-point functions. This would introduce a new IRE $\ka_D$, whose only possible value turned out to be zero, i.~e. no cancelations take place.

We found one solution that features a suppression of the contributions from mixed propagators compared to the other diagrams, i.~e. all diagrams with mixing
propagators can be neglected in the IR. This leads to a system of DSE that is in the IR exactly the same as in standard Landau gauge. As a consequence the calculated value
for the IRE parameter $\ka:=\ka_V=\ka_c$ is also the same, 
$\ka = (93-\sqrt{1201})/98 \approx 0.595353$. Using this result in the DSE of the mixed two-point function its IRE can be calculated. However, we did not find a unique solution but several allowed values. The lowest values are $0.0669$ and $0.9814$. For all values the $AV$-propagator is IR suppressed. An important feature of this solution is that the DSEs reduce to the same system as for standard Landau gauge. There is much information available on that solution that can be used here, e.~g. the use of a bare Faddeev-Popov ghost-gluon vertex is known to be sufficient in numerical calculations with no immediate need to improve on the kinematic dependence. 

We found a second possible solution, where all IREs are strictly related by a scaling relation, $\ka_A+2\ka_c=\ka_A+2\ka_V=\ka_V+2\ka_{AV}=0$, but the inherent structure of the equations did not allow us to find a value for the IRE parameter $\ka$. Nevertheless we can infer from the non-negativity of $\ka_c$ that the qualitative behavior of this solution is the same as that of the first one. In this case the relations between the IREs of the two-point functions and the propagators are just the negative of each other, e.~g. $\ka_A=-\de_A$.
Both solutions found correspond to the same qualitative IR behavior of the propagators: The gluon propagator is IR suppressed, the ghost and auxiliary field propagators are IR enhanced with the same IRE and the mixed propagator is IR suppressed. In case II this suppression is more pronounced.

In obtaining 
our results we had to improve the usual power counting, which normally does not need to differ between the tensors of Green functions. Because of the mixing,
however, the corresponding dressing functions can combine in different ways for different tensors. Thus it was necessary to specify a concrete basis. 
Choosing the tensors that appear in the Lagrangian leads already to a $VV$-propagator with two different tensors. The (possibly) different behavior of these two tensors was important in determining the 
solutions presented here.

In the present article we only worked with the conventional Gribov-Zwanziger action \cite{Zwanziger:1989mf}. As this requires that the horizon condition is fulfilled the propagators of the Faddeev-Popov ghost and the auxiliary fields $\omega$ and $U$ are necessarily divergent \cite{Zwanziger:2009je}. In functional equations this corresponds to a possible choice for the boundary condition of the ghost two-point DSE \cite{Fischer:2008uz}. A different choice leads to the decoupling solution and of course the question arises if this solution could be realized somehow within the Gribov-Zwanziger scenario. It is worth reminding one here that a refinement of the Gribov-Zwanziger theory \cite{Dudal:2008sp} provides in fact a way  to get a finite non-vanishing gluon propagator and a finite ghost dressing function by introducing condensates of the auxiliary fields. Thus,  a direct connection between the refined Gribov-Zwanziger scenario and the corresponding boundary condition in functional equations is to be expected. Zwanziger's argument about cutting off the functional integral at the Gribov horizon is not invalidated by imposing the decoupling boundary condition. However, an interesting difference between the Gribov-Zwanziger and the refined Gribov-Zwanziger scenarios is that in the latter no longer the configurations directly at the horizon dominate but configurations within the Gribov region. This can be seen by calculating the ghost self-energy $\sigma(k^2)$ \cite{Dudal:2008sp}. The study of the refined Gribov-Zwanziger theory and of the related decoupling solution within the functional set up presented here is under  investigation.

To continue and complete this work it will be interesting to try a complete numerical solution of the equations. The information obtained here will definitely be important input for this, as it is now known which diagrams are IR leading and which are the relevant tensors. Even without the full solution available we can conclude that taking into account the restriction to the Gribov region in the path integral does not affect the IR properties as obtained from functional equations: The scaling relation remains valid and the auxiliary fields behave as the Faddeev-Popov ghost.

\begin{acknowledgments}
We thank Daniel Zwanziger for valuable discussions and Axel Maas for a critical reading of the manuscript and helpful discussions.
M.~Q.~H. is supported by the Doktoratskolleg ``Hadrons in Vacuum, Nuclei and Stars'' of the FWF under contract W1203-N08. 
Furthermore he is indebted to the Instituto de F\'isica  of the Universidade do Estado do Rio de Janeiro, where this work was carried out.
R.~A. acknowledges the support of the European Community-Research Infrastructure
Integrating Activity ``Study of Strongly Interacting Matter'' (acronym HadronPhysics2, Grant Agreement n. 227431)
under the Seventh Framework Programme of the EU.
S.~P.~S gratefully acknowledges the Conselho Nacional de Desenvolvimento Cient\'ifico e Tecnol\'ogico (CNPq-Brazil), the Faperj, Funda\c{c}\~{a}o de
Amparo \`{a} Pesquisa do Estado do Rio de Janeiro, the Latin American Center for Physics (CLAF), the SR2-UERJ and the
Coordena\c{c}\~ao de Aperfei\c{c}oamento de Pessoal de N\'ivel Superior (CAPES) for financial.

\end{acknowledgments}

\appendix

\section{Extension of power counting to mixed propagators}
\label{sec:general-IRE}

We present in this appendix the calculations to
obtain eqs. (\ref{eq:lowerBoundMod}) and (\ref{eq:leadingDiagramMod}). They are extensions of calculations given in ref. \cite{Huber:2009wh} to the case of mixed propagators. For the convenience of the reader we also give an example of how to get the inequalities (\ref{eq:props-vert-ineq}) and (\ref{eq:props-ineq}).

\subsection{IRE for an arbitrary diagram and lower bound on the IRE}

Eq. (\ref{eq:lowerBoundMod}) is the formula for the IRE of an arbitrary diagram. It can be obtained by counting all the IREs of propagators and dressed vertices:
\begin{align}\label{eq:deltav-start}
\ka_{v}= & l\frac{d}{2}+\sum_{i}n_i(\delta_{{i}}-1)+\sum_{vertices,r\geq3}n^{d}_{{i_1}\ldots {i_r}}(\ka_{{i_1}\ldots {i_r}}+c_{{i_1}\ldots {i_r}})+\nnnl
 & +\sum_{vertices,r\geq3}n^{b}_{{i_1}\ldots {i_r}}c_{{i_1}\ldots {i_r}}-c_{v}.
\end{align}
$n_i$ are the number of internal propagators with IR exponents $\delta_{{i}}$, whereas the numbers of vertices are $n_{{i_1}\ldots {i_r}}$. Superscripts $d$ and $b$ stand for dressed and bare, respectively. In case none is given, we refer to both. The sums $\sum_{vertices,r\geq3}$ extend over all vertices with at most $r$ legs. $m_{i}$ is the number of external legs.

Without mixed propagators it is possible to use topological relations to get rid of the internal propagators. Here, however, the dependence on the mixed propagators will remain. To avoid cumbersome notation we restrict ourselves now to the case that we only have $A$- and $V$-fields. The inclusion of the fermionic ghost field goes along the lines as presented in ref. \cite{Huber:2009wh}.
The first relation expresses the number of loops in terms of the numbers of propagators and vertices:
\begin{align}\label{eq:numberOfLoops}
l=\sum_{i=A,V}n_i+n_{AV}+1-\sum_{vertices,r\geq3}n_{{i_1}\ldots {i_r}}.
\end{align}
Furthermore we express the number of internal $A$- and $V$-propagators by
\begin{align}\label{eq:relation-verts-props}
n_i=\frac{1}{2}\left(\sum_{vertices,r\geq3}k_{{i}}^{{i_1}\ldots {i_r}}n_{{i_1}\ldots {i_r}}-m_{i}-n_{AV}\right), \quad \quad i=A,V,
\end{align}
where $k_{{i}}^{{i_1}\ldots {i_r}}$ denotes the number of times the field $\phi_{i}$ appears in the vertex $\phi_{i_1}\ldots \phi_{i_r}$.
Plugging these expressions into \eref{eq:deltav-start}, we get
\begin{align}
\ka_{v} = & \left(\sum_{i=A,V}\frac{1}{2}\left(\sum_{vertices,r\geq3}k_{{i}}^{{i_1}\ldots {i_r}}n_{{i_1}\ldots {i_r}}-m_{i}-n_{AV}\right)+1+n_{AV}-\sum_{vertices,r\geq3}n_{{i_1}\ldots {i_r}}\right)\frac{d}{2}+n_{AV}(\de_{AV}-1)+\nnnl
 & +\sum_{i=A,V}\frac{1}{2}\left(\sum_{vertices,r\geq3}k_{{i}}^{{i_1}\ldots {i_r}}n_{{i_1}\ldots {i_r}}-m_{i}-n_{AV}\right)(\delta_{{i}}-1)+\nnnl
 & +\sum_{vertices,r\geq3}n^{d}_{{i_1}\ldots {i_r}}\left(\ka_{{i_1}\ldots {i_r}}+2-\frac{r}{2}\right)+\sum_{vertices,r\geq3}n^{b}_{{i_1}\ldots {i_r}}\left(2-\frac{r}{2}\right)-2+\frac{1}{2}\sum_{i}m_{i}=\nnnl
= & \left(\frac{d}{2}-2\right)\left(1-\frac{1}{2}\sum_{i=A,V}m_{i}\right)-\frac{1}{2}\sum_{i}m_{i}\delta_{{i}}+
 \sum_{vertices,r\geq3}n^{d}_{{i_1}\ldots {i_r}}\left(-\frac{d}{2}+\ka_{{i_1}\ldots {i_r}}+2-\frac{r}{2}\right)+\nnnl
 & +\sum_{i=A,V}\frac{1}{2}\left(\sum_{vertices,r\geq3}k_{{i}}^{{i_1}\ldots {i_r}}n_{{i_1}\ldots {i_r}}(\frac{d}{2}+\delta_{{i}}-1)\right)+
 \sum_{vertices,r\geq3}n^{b}_{{i_1}\ldots {i_r}}\left(-\frac{d}{2}+2-\frac{r}{2}\right)+\nnnl
 & +n_{AV}\left((\delta_{AV}-1)-\frac1{2}\sum_{i=A,V}(\de_{i}-1)\right).
\end{align}

This leads to
\begin{align}\label{eq:IREArbitraryDiagram}
\ka_{v} = & \left(\frac{d}{2}-2\right)\left(1-\frac{1}{2}\sum_{i}m_{i}\right)-\frac{1}{2}\sum_{i}m_{i}\delta_{{i}}+\frac1{2}n_{AV}(2\de_{AV}-\de_A-\de_V)+\nnnl
 & +\sum_{vertices,r\geq3}n^{d}_{{i_1}\ldots {i_r}}\left(\left(\frac{d}{4}-1\right)(r-2)+\ka_{{i_1}\ldots {i_r}}+\frac{1}{2}\sum_{i}k_{{i}}^{{i_1}\ldots {i_r}}\delta_{{i}}\right)+\nnnl
 & +\sum_{vertices,r\geq3}n^{b}_{{i_1}\ldots {i_r}}\left(\left(\frac{d}{4}-1\right)(r-2)+\frac{1}{2}\sum_{i}k_{{i}}^{{i_1}\ldots {i_r}}\delta_{{i}}\right).
\end{align}
In ref. \cite{Huber:2009wh} it was shown that the last two terms are non-negative. We present one explicit example below. This conclusion does not change for mixed propagators. So the lower bound on the IRE of a diagram $v$ is
\begin{align}
\ka_{v,max} = & \left(\frac{d}{2}-2\right)\left(1-\frac{1}{2}\sum_{i}m_{i}\right)-\frac{1}{2}\sum_{i}m_{i}\delta_{{i}}+\frac1{2}n_{AV}(2\de_{AV}-\de_A-\de_V).
\end{align}
In contrast to the non-mixing case it depends on the internal propagators, but only on the mixed ones. The value of $2\de_{AV}-\de_A-\de_V$ has to be considered for each case. It should be non-negative, because otherwise the IRE is unbounded from below, since there are diagrams with an arbitrary number of mixed propagators. They can be explicitly constructed by appropriately replacing dressed quantities by their respective DSEs.

\subsection{Non-negativity of combinations of IREs}
\label{subsec:non-neg}

For the IR analysis done in this article we also relied on inequalities obtained from RGEs. These functional equations also form an infinitely large tower of equations and are derived from the effective average action, see for example the reviews \cite{Berges:2000ew,Pawlowski:2005xe}. The average effective $\Gamma^k$ depends on the momentum scale $k$, above which all fluctuations are integrated out. The limits $k\rightarrow 0$ and $k\rightarrow \infty$ correspond to the standard effective and classical actions, respectively. The dependence of the average effective action on the scale $k$ is given by:
\begin{align}\label{eq:av-eff-action}
 k\, \partial_k \Gamma^k[\phi]=& \frac1{2} k\, \partial_k R^k_{ij} G_{ij}.
\end{align}
where $\partial_k:=\partial/\partial k$ and the indices $i$ and $j$ represent field type, momentum and internal indices like Lorentz or color indices. Summation/Integration over indices appearing twice is understood. The connected two-point function is
\begin{align}
 G_{ij}=\frac1{\Gamma^k_{ij}[\Phi] + R^k_{ij}}
\end{align}
and the quantity $R_{ij}$ is a regulator that implements the momentum cutoff at the scale $k$. $\Gamma^k_{ij}[\Phi]$ is the corresponding two-point function.

RGEs are derived by functional differentiation of \eref{eq:av-eff-action} with respect to fields.
In contrast to DSEs they only feature one-loop equations, all quantities are dressed and  a regulator insertion appears. Forms of regulators especially suited for an IR analysis are discussed in ref. \cite{Fischer:2006vf}.

\begin{figure}[th]
\begin{minipage}{0.48\textwidth}
\includegraphics[width=10cm]{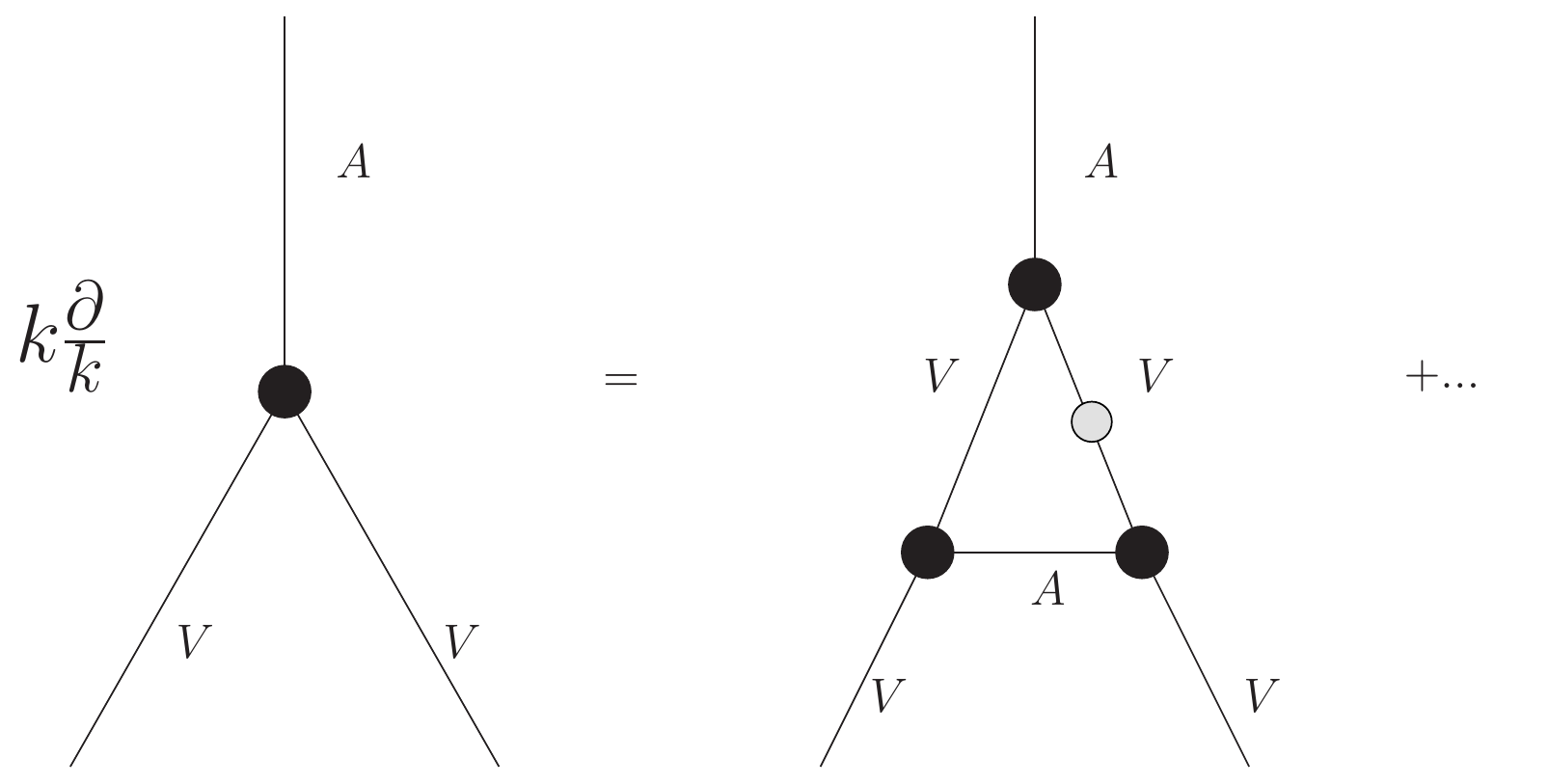}
\caption{\label{fig:RG-3-point}One specific diagram in the RGE of the $AVV$-vertex. Internal lines represent dressed propagators, black blobs dressed vertices. The gray blob is a regulator.}
\end{minipage}
\end{figure}

The tower of RGEs allows one to derive the specific group of inequalities given by
\begin{align} \label{eq:props-vert-ineq-App}
\ka_{{i_1}\ldots {i_r}}+\frac{1}{2}\sum_{i}k_{{i}}^{{i_1}\ldots {i_r}}\delta_{{i}} \geq 0.
\end{align} 
We only provide a simple example here and refer for a complete proof to ref. \cite{Huber:2009wh}.
The simplest case is that of a three-point function. Let us therefore examine for concreteness the $AVV$-vertex. Part of its RGE is depicted in \fref{fig:RG-3-point}.
Power counting yields
\begin{align}
\ka_{AVV}\leq& 3\ka_{AVV}+2\de_V+\de_A,\nnnl
0\leq& \ka_{AVV}+\frac1{2}\left(2\de_V+\de_A\right).
\end{align}
A corresponding analysis of its DSE would yield a similar inequality but with different, less restrictive numerical coefficients. Similar equations can be obtained for all vertices, as can be shown by an iterative proof \cite{Huber:2009wh}, and the set of inequalities can be written down in closed form, see \eref{eq:props-vert-ineq-App}. One should note that these inequalities do not depend on the IREs of two-point functions and thus do not suffer from complications due to the mixing. Furthermore, we do not get additional inequalities if the fields mix.

Another set of inequalities, given by 
\begin{align} \label{eq:props-ineq-App}
\frac1{2} \sum_{i} k_{i}^{{i_1} \ldots {i_r}} \de_{i} \geq 0,
\end{align}
can be derived by taking into account that the IRE of a vertex that appears in the action is non-positive, as can be inferred from its DSE. In the present example this means
\begin{align}
0\leq-\ka_{AVV}\leq& \frac1{2}\left(2\de_V+\de_A\right) \quad \Rightarrow \quad 2\de_V+\de_A\geq0.
\end{align}

One should note that the number of inequalities in the second group, \eref{eq:props-vert-ineq-App}, corresponds to the number of interactions in the Lagrangian, while the first group, \eref{eq:props-ineq-App}, is infinitely large.
Furthermore, the interplay between DSEs and RGEs is nicely exhibited in the derivations of eqs. (\ref{eq:props-vert-ineq-App}) and (\ref{eq:props-ineq-App}). The former can only be derived from the RGEs, while the latter require information provided by the DSEs. Yet, both inequalities are required for the IR analysis.

\subsection{Analysis of leading diagrams in two-point DSEs}

For an arbitrary diagram in a propagator DSE we have from \eref{eq:IREArbitraryDiagram}
\ali{\label{eq:leading-diagram}
\ka_{i} = & \left(\frac{d}{2}-2\right)\left(1-\frac{1}{2}\sum_{j}m_{j}\right)-\frac{1}{2}\sum_{i}m_{i}\delta_{{i}}+\frac1{2}n_{AV}(2\de_{AV}-\de_A-\de_V)+\nnnl
 & +\sum_{\substack{dressed\\vertices}}n^{d}_{{i_1}\ldots {i_r}}\ka_{{i_1}\ldots {i_r}}
  +\sum_{\substack{all\\vertices}}n_{{i_1}\ldots {i_r}}\left(\left(\frac{d}{4}-1\right)(r-2)+\frac{1}{2}\sum_{j}k_{{j}}^{{i_1}\ldots {i_r}}\delta_{{i}}\right)
}
if it is assumed to be part of the leading terms. Here $\ka_i$ can be any two-point IRE and $\de_j$ is restricted to IREs of non-mixing propagators.
Using $\sum_i m_i=2$ in the case of propagators, this can be written as
\ali{
\ka_{i} +\frac{1}{2}\sum_{j}m_{j}\delta_{{i}}-\frac1{2}n_{AV}(2\de_{AV}-\de_A-\de_V)
  -\sum_{\substack{all\\vertices}}n_{{i_1}\ldots {i_r}}\left(\left(\frac{d}{4}-1\right)(r-2)+\frac{1}{2}\sum_{j}k_{{j}}^{{i_1}\ldots {i_r}}\delta_{{i}}\right)=\sum_{\substack{dressed\\vertices}}n^{d}_{{i_1}\ldots {i_r}}\ka_{{i_1}\ldots {i_r}}.
}
We can use the lower bound for the IREs of the vertices on the right-hand side given by the maximally IR divergent solution, \eref{eq:lowerBoundMod}, to get a new inequality:
\ali{
\ka_i+\frac1{2}\sum_j \de_j m_j&-\frac{1}{2}n_{AV}(2\de_{AV}-\de_A-\de_V)-\sum_{\substack{all\\vertices}}n_{{i_1}\ldots {i_r}}\left(\left(\frac{d}{4}-1\right)(r-2)+\frac{1}{2}\sum_{j}k_{{j}}^{{i_1}\ldots {i_r}}\delta_{{i}}\right) \geq \nnnl
\geq& \sum_{\substack{dressed\\vertices}} n^d_{{i_1} \ldots {i_r}}  \left(\left(\frac{d}{2}-2\right)\left(1-\frac{1}{2}\sum_{j}k_{j}^{{i_1} \ldots {i_r}}\right)-\frac1{2} \sum_j \de_{j} k_{j}^{{i_1} \ldots {i_r}} +\frac1{2} \bar{k}^{{i_1} \ldots {i_r}}_{AV}(2\de_{AV}-\de_A-\de_V) \right) .
}
Here $\bar{k}^{{i_1} \ldots {i_r}}_{AV}$ indicates the number of times a mixed propagator is contained in the diagram that determines the IRE of the vertex $\phi_{i_1} \ldots \phi_{i_r}$. Note that $\bar{k}^{{i_1} \ldots {i_r}}_{AV}$ only is different from zero for vertices that necessarily contain an $AV$-propagator like the $AAV$-vertex.
The right-hand side depends on dressed vertices only, indicated by the $d$ superscript of $n$. On the other hand, the left-hand side sums over dressed and bare vertices, so that in total only the bare vertex remains in the sums over vertices:
\ali{
\ka_i+\frac1{2}\sum_j \de_j m_j-n^b_{{i_1}\ldots {i_r}}\left(\left(\frac{d}{4}-1\right)(r-2)+\frac{1}{2}\sum_{j}k_{{j}}^{{i_1}\ldots {i_r}}\delta_{{i}}\right)  -\frac1{2}(2\de_{AV}-\de_A-\de_V)\left( n_{AV}+\sum_{\substack{dressed\\vertices}} n^d_{{i_1} \ldots {i_r}} \bar{k}^{{i_1} \ldots {i_r}}_{AV} \right) \geq0.
}
$\sum_i k_i^{i_1 \ldots i_r}=r$ was used here.

\newpage

\bibliographystyle{utphys}
\bibliography{literature}

\end{document}